\documentclass{aa}  

\usepackage{graphicx}
\usepackage{txfonts}
\usepackage[colorlinks=true, allcolors=blue]{hyperref}
\usepackage{orcidlink}
\usepackage[export]{adjustbox}
\usepackage{multicol}
\usepackage{multirow}
\usepackage{enumitem} 
\usepackage{cuted} 
\usepackage{tikz}

\begin{document} 

   \title{X-ray polarization of reflected thermal emission}

   \titlerunning{X-ray polarization of reflected thermal emission}

   \author{ J.~Podgorn{\'y} \inst{1}\thanks{E-mail: jakub.podgorny@asu.cas.cz}\orcidlink{0000-0001-5418-291X}, M.~Dov{\v{c}}iak \inst{1}\orcidlink{0000-0003-0079-1239},
R.~Goosmann\inst{2},
F.~Marin\inst{2}\orcidlink{0000-0003-4952-0835},
L.~Marra\inst{3,4}\orcidlink{0009-0001-4644-194X},
G.~Matt\inst{5}\orcidlink{0000-0002-2152-0916},
A.~R\'o\.za\'nska\inst{6}\orcidlink{0000-0002-5275-4096}, R.~Taverna\inst{4}\orcidlink{0000-0002-1768-618X}, \and M.~Gupta\inst{1}\orcidlink{0000-0003-0976-8932}
   }

   \institute{
   Astronomical Institute of the Czech Academy of Sciences, Bo\v{c}n\'{i} II 1401/1, 14100 Praha 4, Czech Republic
   \and Universit\'{e} de Strasbourg, CNRS, Observatoire Astronomique de Strasbourg, UMR 7550, 67000 Strasbourg, France
   \and INAF Istituto di Astrofisica e Planetologia Spaziali, Via del Fosso del Cavaliere 100, 00133 Roma, Italy
   \and Dipartimento di Fisica e Astronomia, Universit\`{a} degli Studi di Padova, Via Marzolo 8, 35131 Padova, Italy
   \and Dipartimento di Matematica e Fisica, Università degli Studi Roma Tre, Via della Vasca Navale 84, 00146 Roma, Italy
   \and Nicolaus Copernicus Astronomical Center, Polish Academy of Sciences, Bartycka 18, 00-716 Warsaw, Poland
   }

  \authorrunning{J. Podgorn{\'y} et al.}

   \date{Received ...; Accepted ...}

 
\abstract
{X-ray thermal emission is inherent in neutron-star and black-hole X-ray binary systems. Within these systems, it may reflect from optically thick matter, which will create characteristic observable X-ray spectro-polarimetric features. We compute rest-frame reflection spectra and the corresponding energy-dependent linear polarization degree and angle for (un)polarized single-temperature blackbody spectra impinging on a partially ionized constant-density optically thick slab. We use a combination of a Monte Carlo simulation that takes into account scattering, absorption, and spectral lines, with a non-LTE radiative transfer pre-computation of the ionization structure of the slab in photo-ionization equilibrium. We discuss the impact of the reflector's ionization and of the incident spectral shape on the obtained energy dependence of polarization. For isotropic incident intensity and for rather parallel than perpendicular unidirectional illumination with respect to the slab normal direction, despite the presence of highly polarized absorption features and low-polarized spectral lines, an underlying scattering-induced increase of polarization degree with energy in mid to hard X-rays naturally arises due to multiple Compton-scattering energy shifts. Such re-processing effect is particularly apparent in 2--8 keV for steep incident X-ray spectra reflecting from highly-ionized optically thick media. Integration of the resulting local reflection tables in specific large-scale reflection geometries occurring in X-ray binary systems, including relativistic effects, will be presented in a follow-up paper. Nonetheless, we anticipate that the obtained local energy-dependent features will imprint at large distances from the source to the observed X-ray polarization, and could contribute to the observed increase of total polarization degree with energy in 2--8 keV in some accreting systems by the IXPE mission.
   }


   \maketitle
%

\section{Introduction}
\label{sec:intro}

X-ray emission from accreting neutron stars (NSs) and black holes (BHs) in X-ray binary systems (XRBs) is dominated by a soft thermal multi-temperature blackbody component from the inner accretion disc (or the surface or the boundary layer of the NS), and a harder non-thermal Comptonized power-law tail from a hot corona \citep{Shakura1973,Sunyaev1980,Mitsuda1984}. This dual-component nature has been fundamental in shaping modern phenomenological models for XRBs and linking them to the theory of accreting compact objects \citep{Zdziarski2004, Done2007}.

When the thermal photons from the disc or the NS surface or boundary layer strike optically-thick, partially-ionized matter—such as the disc atmosphere or optically-thick equatorial outflows—they become partially reflected, producing characteristic X-ray reflection features (e.g. iron K$\alpha$ line, Compton hump) and a secondary re-processed blackbody spectrum from thermalization, as shown by \citet{Ballantyne2004, Ballantyne2004b} based on \citet{Ross1993,Ross2005}. These works formed a foundation of the {\tt BBrefl} and {\tt reflionxBB} rest-frame reflection models for a constant-density slab in photo-ionization equilibrium. The authors emphasized that blackbody-illuminated reflection spectra can be directly observed during X-ray bursts and the so-called high-soft XRB accretion states. Later computational efforts, including the widely-used {\tt xillver} (rest-frame) and {\tt relxill} (relativistic) families of fitting tools \citep{Garcia2010,Dauser2014, Garcia2014}, provided angle-dependent, ionization- and density-resolved models, applied extensively to accreting BHs, assuming an incident power-law spectral shape. Building on the model suite, \citet{Garcia2022} introduced new reflection models—{\tt xillverNS} (rest-frame) and {\tt relxillNS} (relativistic)—specifically tailored for NS XRBs, in which the primary continuum is described by a Planck function instead, modifying the reflection spectrum and its dependence on ionization state and disc density. These models have become useful for interpreting high-resolution X-ray spectral data from XRBs \citep{Ludlam2020,Ludlam2022, Ursini2023c}.
\\ \indent
Linear polarization degree and angle are now directly observable with high precision in mid X-rays thanks to the launch of the Imaging X-ray Polarimetry Explorer \citep[IXPE,][]{Weisskopf2022}, operating in the 2--8 keV band. The acquired polarization data by IXPE are essential for XRB studies \citep[see reviews by][for the results from the first 2.5 years of the mission's operation]{Dovciak2024, Poutanen2024, Ursini2024}. In numerous XRB sources observed by IXPE, a significant reflection fraction of the total flux in the 2--8 keV band was spectroscopically detected, which calls for the development of accurate spectro-polarimetric reflection models that would take into account the competing scattering, absorption, and spectral-line effects for diverse sources, reflectors, and geometries. A multi-component analysis incorporating detailed energy-dependent predictions for the reflection component is a key interpretation tool for the current and forthcoming X-ray spectro-polarimetric campaigns.

Classical single-scattering and multiple-scattering reflection polarization estimates are provided in \cite{Chandrasekhar1960} for a semi-infinite electron-scattering atmosphere. X-ray polarization of reflection from a neutral slab was further examined by \cite{Matt1993b} and \cite{Poutanen1996b} with the effects of Compton scattering, absorption, and spectral lines. The 0.1--100 keV reflection-induced polarization from intermediately (partially) ionized optically thick constant-density slabs in photo-ionization equilibrium was calculated by \citet{Podgorny2022} for a locally impinging X-ray power-law spectrum. The study introduced the rest-frame reflection results in a tabulated format—to be used by (relativistic) integration codes for specific geometries \citep{Podgorny2023a, Podgorny2024}. Simultaneous spectro-polarimetric modeling was achieved via combined Monte Carlo (MC) simulations, using the {\tt STOKES} code \citep{Goosmann_2007, Marin_2012, Marin_2015, Marin_2018_UV}, with non-LTE ionization structure pre-computations in photo-ionization equilibrium, using the {\tt TITAN} code \citep{Dumont2003}. The same approach with {\tt STOKES} tied to an iterative radiative transfer solver was also adopted for calculation of the transmission of thermal radiation through a partially-ionized disc atmosphere, using independently the {\tt TITAN} and {\tt CLOUDY} \citep{Ferland2013,Ferland2017} codes for ionization structure pre-computations \citep{Taverna2021, Ratheesh2024, Marra2026}, resulting in a reasonable agreement between the two.

In this work, we use the same method as in \citet{Podgorny2022} to simulate rest-frame 2--80 keV spectro-polarimetric properties of the radiation reflected from a partially-ionized, optically-thick, constant-density slab illuminated by a single-temperature blackbody spectrum, instead of a power-law spectral shape. The model is presented in Section \ref{sec:model}. In Section \ref{sec:results}, the results are presented concerning the impact of the incident spectral shape and slab ionization. We discuss the model assumptions and compare the obtained {\tt STOKES} spectra to the {\tt xillverNS} spectra in Section \ref{sec:discussion}, which enables us to discuss the role of a thermalized reflection sub-component and to estimate the reflection-induced polarization in the soft X-rays. We conclude in Section \ref{sec:conclusion}. In Appendix \ref{cross-validation} we provide additional consistency checks of the {\tt STOKES} code in X-rays with other MC codes and (semi-)analytical results from literature.

\section{Model}
\label{sec:model}

\subsection{{\tt TITAN} code}

To model the ionization equilibrium of the slab illuminated from one side by external X-rays, we use the X-ray photo-ionization code {\tt TITAN} in the same way as in \citet{Podgorny2022}, where it was solved for an impinging power-law photon flux $N_\mathrm{E}\sim E^{-\Gamma}$ with a photon-index $\Gamma$. In this work, the incident photon flux is a Planck function $N_\mathrm{E}\sim E^2  [\exp{(\frac{E}{k_\mathrm{B}T})}-1]^{-1}$ dependent on the blackbody temperature $T$, $k_\mathrm{B}$ being the Boltzmann constant. {\tt TITAN} solves non-LTE radiative transfer in plane-parallel geometry coupled to the ionization and energy balance equations \citep{Dumont2003}. Starting from incident radiation, {\tt TITAN} iteratively computes at each depth the layer’s temperature, electron density, and ion abundances for the lines and continuum up to 26 keV, which is enough for the required output on ionization properties. We assume a constant neutral hydrogen density $n_\mathrm{H} = 10^{15} \, \textrm{cm}^{-3}$ throughout the slab and a solar abundance of elements from \citet{Asplund2005} with metallicity $A_\mathrm{Fe} = 1.0$. The irradiation source is located directly on top of the slab, and we select an ionization parameter $\xi$ defined as \citep{Tarter1969}
\begin{equation}\label{xi}
	\xi = \dfrac{4 \pi \int F_\mathrm{E} (E;k_\mathrm{B}T) \, \mathrm{d}E}{n_\mathrm{H}} \textrm{ ,}
\end{equation}
where $F_\mathrm{E} = E N_\mathrm{E}$ is the radiation energy flux received on the surface of the slab. For the pre-computation of the ionization structure, we use angle-averaged illumination of the slab. Numerically, {\tt TITAN} uses the Accelerated Lambda Iteration (ALI) method \citep{Cannon1973, Scharmer1981, Olson1986, Hubeny2003}. We obtain the physical conditions within the slab stratified in $\sim 500$ layers up to an ending Thomson optical depth $\tau \sim 7$, which is used as input for the subsequent MC spectro-polarimetric radiative transfer modeling. For the conversion of the pre-computed ionization structure to {\tt STOKES} we use a specific C++ script that averages the output from {\tt TITAN} to $\sim 50$ layers, forming the adopted vertical stratification of the optically thick reflecting medium.

\subsection{{\tt STOKES} code}

\begin{figure} 
\centering
\includegraphics[width=1\linewidth, trim={5.25cm 5cm 11.3cm 4cm}, clip]
{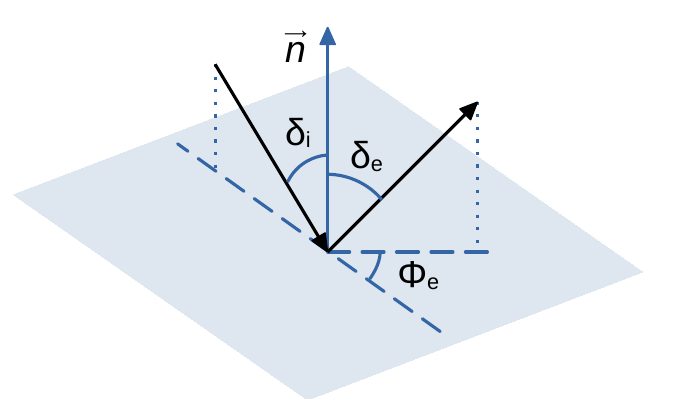}
\caption{Geometry of the rest-frame reflection. Three angles $\delta_\mathrm{i}$, $\delta_\mathrm{e}$, and $\Phi_\mathrm{e}$ are defined by the direction of the incident and emission vectors (black arrows) with respect to the slab normal and with respect to each other when projected to the slab plane.} \label{fig:sketch}
\end{figure}

Taking into account the slab’s structure from {\tt TITAN}, the {\tt STOKES} code performs an MC radiative transfer to generate the emergent spectrum and polarization simultaneously in the same physical conditions and plane-parallel geometry as in {\tt TITAN} per each point in the selected grid (see Section \ref{tables}). The {\tt STOKES} code is developed specifically for polarization tracking, and it registers individual photons undergoing photo-electric and free-free absorption, resonant and fluorescent line emission, and Compton down-scattering. For the main computational results presented in this work, we use version v2.33, which is suitable for single-temperature blackbody sources and which is optimized for X-rays. It computes all four Stokes parameters ($I$, $Q$, $U$, $V$) emergent in the local frame, although here we focus only on the $I$, $Q$, and $U$ to describe the linear polarization in accordance with our observational motivations. {\tt STOKES} returns energy-resolved spectro-polarimetric results for a range of incident/emergent angles and incident polarization states. Inside the MC simulation we set the locally incident inclination angle to the slab normal $\delta_\mathrm{i} = \arccos{(\mu_\mathrm{i})}$, locally emergent inclination angle to the slab normal $\delta_\mathrm{e} = \arccos{(\mu_\mathrm{e})}$, and the locally emergent azimuthal angle $\Phi_\mathrm{e}$ (for a sketch see Figure \ref{fig:sketch}). We choose the same sharp low- and high-energy cut-offs for the incident spectrum as in \citet{Podgorny2022}: $E_\mathrm{l} = 10^{-1.1} \, \textrm{keV}$ and $E_\mathrm{h} = 10^{2.4} \, \textrm{keV}$, respectively, similarly to the works of \cite{Ballantyne2004} and \cite{Garcia2022}.

\subsection{Tabulated grid of computations}\label{tables}

The resulting Stokes parameters $I$, $Q$, and $U$ from the {\tt STOKES} code are tabulated\footnote{The final blackbody local reflection tables under different flavors are added to the power-law local reflection tables, available at \url{https://github.com/jpodgorny/stokesBB_tables}, including documentation for direct usage inside {\tt XSPEC} \citep{Arnaud1996}.} in the {\tt FITS} files format, conforming to the {\tt OGIP} standard \citep{FITS, OGIP}. We use the energy resolution of 160 logarithmically spaced bins between 2 and 80 keV, which is a compromise between reasonable computational times\footnote{For each point in the final grid, we used at least $10^9$ simulated photons for sufficient suppression of the MC numerical noise, while some corners of the parameter space required more photons. The entire grid took about 2 months to compute, using a typical capacity of an institutional computer cluster.} and a resolution allowing inspection of the main X-ray spectral and polarization features. We emphasize that the tabular model is not intended for high-resolution spectral fitting. It should serve instead for energy-dependent estimation of X-ray polarization properties in the mid and hard X-rays, where the adopted energy resolution is far beyond the capabilities of contemporary X-ray polarimeters, such as the 2--8 keV IXPE mission or the 15--80 keV XL-Calibur balloon experiment \citep{Abarr2021}. The final model parameters are $\xi$, $k_\mathrm{B}T, \mu_\mathrm{i}, \mu_\mathrm{e}, \Phi_\mathrm{e}$, and the incident polarization state, for which we choose three independent cases as in \citet{Podgorny2022}, allowing interpolation for arbitrary incident polarization. The resulting tables are summarized in Table \ref{FITSgrid}. 
\begin{table}
 \footnotesize 
	\centering
	\caption{Description of the local reflection tables computed by {\tt STOKES} that are attached in the {\tt FITS} format.}
	\begin{tabular}{ll}
		Number of tables: & 3 -- 100\% vertical, 100\% diagonal, \\ &\,\,\,\, and no incident polarization \\
        Stokes parameters: &$I$, $Q$, $U$, normalized according to (\ref{norm1})  \\
		Spectral units:            & $10^{20}\times[\textrm{cts} \,\,\, \textrm{cm}^{-2} \,\,\,  \textrm{s}^{-1}]$        \\
		Energy range:     & $2 \textrm{ keV}$ to $80 \textrm{ keV}$                       \\
		Energy binning:      & 160 bins with $\Delta \log E = 0.01$                                             \\
		$\xi$ $[\textrm{erg} \,\,\,  \textrm{cm} \,\,\,  \textrm{s}^{-1}]$            & {\{}$10,20,50,100,200,500,1000,2000,$ \\ & \,\,\,\,$5000,10000,20000${\}}       \\
        $k_\mathrm{B}T$ $[\textrm{keV}]$            & {\{}$0.2,0.5,1,2,5,10${\}}                                                             \\
		$\mu_\mathrm{i}$          & {\{}$0.0,0.1,0.2,0.3,0.4,0.5,0.6,0.7,0.8,$ \\ &\,\,\,\, $0.9,1.0${\}}                    \\
		$\mu_\mathrm{e}$            & {\{}$0.025,0.075,0.175,0.275,0.375,0.475,$ \\
		&
		\multicolumn{1}{l}{$\ \ \ 0.575,0.675,0.775,0.875,0.975${\}}}                        \\
		$\Phi_\mathrm{e}$           & {\{}$7.5^{\circ},22.5^{\circ},37.5^{\circ},52.5^{\circ},67.5^{\circ},82.5^{\circ},$                               \\
		&
		\multicolumn{1}{l}{$\ \ \ 97.5^{\circ},112.5^{\circ},127.5^{\circ},142.5^{\circ},157.5^{\circ},$}                                              \\
		& \multicolumn{1}{l}{$\ \ \ 172.5^{\circ},187.5^{\circ},202.5^{\circ},217.5^{\circ},232.5^{\circ},$}                                \\
		& \multicolumn{1}{l}{$\ \ \ 247.5^{\circ},262.5^{\circ},277.5^{\circ},292.5^{\circ},307.5^{\circ},$}         \\
        
		& \multicolumn{1}{l}{$\ \ \ 322.5^{\circ},337.5^{\circ},352.5^{\circ}${\}}}         \\
		Extensions:       & Primary Header  -- description of the tables                                             \\
		& `PARAMETERS’      -- parameter values                                                      \\
		& `ENERGIES’     -- low and high energy \\ & \,\,\,\, bin edges                                      \\
		& `SPECTRA’  -- values of the Stokes \\ & \,\,\,\, parameters and corresponding model \\
		&\,\,\,\,\,parametric values          \\
	\end{tabular}%
	\label{FITSgrid}
\end{table}

To store the Stokes parameters $I$, $Q$, and $U$, generally denoted as $S$, we use the same normalization approach as in \citet{Podgorny2022}, only exchanging the original power-law with a blackbody incident spectral shape. This means that to recover the final spectra in units of $[\textrm{cts} \,\,\,  \textrm{cm}^{-2} \,\,\, \textrm{s}^{-1}]$, the raw output from {\tt STOKES}, which is in $[\textrm{cts} \,\,\,  \textrm{cm}^{-2} \,\,\, \textrm{s}^{-1}]$ per angular bin, needs to be normalized as
\begin{equation}\label{norm1}
	N(E) = \dfrac{\int_{E_\mathrm{l}}^{E_\mathrm{h}} E^2  [\exp{(\frac{E}{k_\mathrm{B}T})}-1]^{-1} \,\mathrm{d}E\,\,  \xi\, n_\mathrm{H}}{\int_{E_\mathrm{l}}^{E_\mathrm{h}} E^3  [\exp{(\frac{E}{k_\mathrm{B}T})}-1]^{-1} \,\mathrm{d}E\,\, 4 \pi}\dfrac{S(E)}{N_\mathrm{tot} \Delta \mu_\mathrm{e} \Delta \Phi_\mathrm{e}} \textrm{ ,}
\end{equation}
where $\Delta\mu_\mathrm{e} = 0.05$ and $\Delta \Phi_\mathrm{e} = 15^\circ$ are the corresponding angular bin sizes and $N_\mathrm{tot}$ is the number of photons launched per simulation. We additionally multiply the spectra by an arbitrary $10^{-20}$ factor for storage facilitation inside {\tt FITS} files.

\section{Results}
\label{sec:results}

We will discuss the polarization results in terms of linear polarization degree, $p$, and the polarization angle, $\Psi$, that are obtained from the Stokes parameters $I$, $Q$, and $U$ through the usual expressions
\begin{equation}\label{ppsidef}
	\begin{aligned}
		p &= \dfrac{\sqrt{Q^2+U^2}}{I} \textrm{ ,} \\
		\Psi &= \dfrac{1}{2}\textrm{\space}\arctan_2\left(\dfrac{U}{Q}\right)   \textrm{ ,}
	\end{aligned}
\end{equation}
where $\arctan_2$ denotes the quadrant-preserving inverse of the tangent function and $\Psi = 0^\circ$ corresponds to a polarization vector oriented along the direction of the local reflecting surface normal, projected into the polarization plane. $\Psi$ increases in the counterclockwise direction from the observer’s point of view. For symmetric situations, we will use the convention of positive and negative polarization degree when $\Psi = 0^\circ$ and $\Psi=90^\circ$, respectively.

The attached re-processing spectro-polarimetric tables include dependencies on incident and emission angles, as well as incident polarization. However, we skip the associated discussion, because the emergent Stokes parameters for blackbody irradiation depend on these model parameters in the same way as for the power law, which was already discussed in \citet{Podgorny2022}. The related discussion is unaffected by the choice of incident spectral shape. We re-created the figures from \citet{Podgorny2022} to prove the point, including comparisons with Chandrasekhar's analytical results \citep{Chandrasekhar1960}, which additionally verifies the new implementation. The key parameters of the new model to examine in this work are
\begin{itemize}[label=\textbullet]
    \item the blackbody temperature $T$, effectively changing the incident spectral steepness, similarly to the $\Gamma$ index in the original power-law model;
    \item the ionization parameter $\xi$, which defines the reflection spectral amplitude and effectively scales the ion abundances in each layer, depending on the incident spectral shape.
\end{itemize}

\begin{figure*} 
\centering
\begin{tikzpicture}[
x=1pt, y=1pt,
inner sep=0pt,
outer sep=0pt]
\node[anchor=south west] (base) at (0,0)
  {\includegraphics[width=\linewidth,     
                   trim={3.6cm 1.2cm 5.0cm 0.3cm},
                   clip]{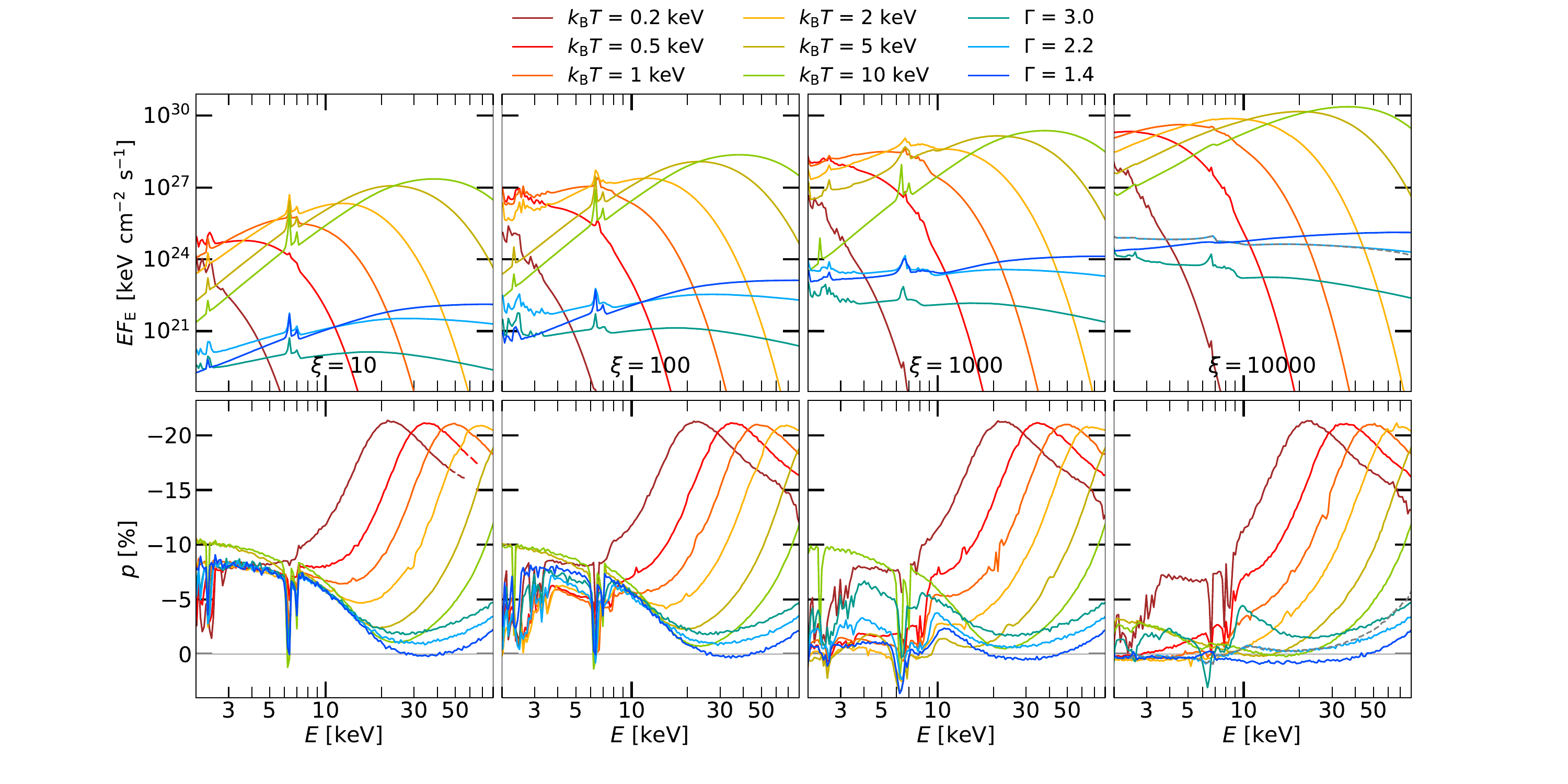}};
\node[anchor=south west] at (35,52)
  {\includegraphics[width=0.095\linewidth]{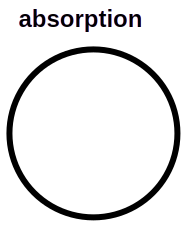}};

\node[anchor=south west] at (171,76)
  {\includegraphics[width=0.09\linewidth]{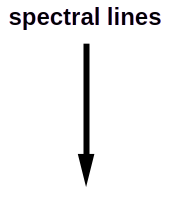}};

\node[anchor=south west] at (79,33)
  {\includegraphics[width=0.12\linewidth]{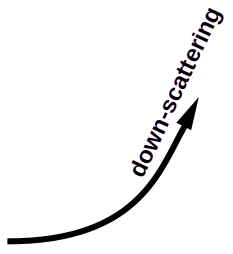}};

\end{tikzpicture}
\caption{Geometry-averaged local reflection spectra, $EF_\mathrm{E}$ (top row), and the corresponding polarization degree, $p$ (bottom row), versus energy in the 2--80 keV range for various incident spectral shapes (in the same color code per panel) and ionization parameters $\xi = 10, 100, 1000, 10000  \,\, \textrm{erg} \,\,\, \textrm{cm} \,\,\, \textrm{s}^{-1}$ (from left to right panel columns). The results are integrated in all incident inclination angles, $\mu_\mathrm{i}$, all emission azimuthal angles, $\Phi_\mathrm{e}$, and plotted for one emission inclination angle, $\mu_\mathrm{e} = 0.475$. We show six values of temperature of the incident unpolarized single-temperature blackbody radiation $k_\mathrm{B}T = 0.2,0.5,1,2,5,10\,\,\textrm{keV}$ and three values of the power-law index of the incident power-law radiation $\Gamma = 1.4,2.2,3.0$, taken from \citet{Podgorny2022} for comparison. The dashed gray line in the rightmost panels ($\xi = 10000  \,\, \textrm{erg} \,\,\, \textrm{cm} \,\,\, \textrm{s}^{-1}$) shows the calculation for $\Gamma = 2.2$ using smaller high-energy cut-off in {\tt STOKES} than for the rest of the computations with $E_\mathrm{h} = 10^{2.4} \, \textrm{keV}$ by a factor of 2.} \label{fig:energy_dep}
\end{figure*}

Figure \ref{fig:energy_dep} shows the resulting spectra and polarization degree for various spectral shapes of unpolarized incident radiation and changing ionization parameter. Apart from various temperatures of the impinging thermal emission, we include three cases of power-law incident spectra from \citet{Podgorny2022} with different $\Gamma$, normalized in the same manner for comparison. The local reflection is integrated in the incident inclination angles with angular distribution as for an incident intensity illumination and uniformly integrated in the emission azimuthal angles, which has a minimal impact on spectra. The amplitude of the obtained polarization fraction is reduced due to the geometrical averaging, but it allows studying the energy profile of polarization with lower MC numerical noise. We show the spectra in $EF_\mathrm{E} = E^2N_\mathrm{E}$, where the angle-averaged photon flux (for all three Stokes parameters) in $[\textrm{cts} \,\,\,  \textrm{cm}^{-2} \,\,\, \textrm{s}^{-1} \,\,\, \textrm{keV}^{-1}]$ is normalized according to
\begin{equation}\label{norm2}
    \begin{split}
	N_\mathrm{E}(E) & = \dfrac{\int_{E_\mathrm{l}}^{E_\mathrm{h}} E^2  [\exp{(\frac{E}{k_\mathrm{B}T})}-1]^{-1} \,\mathrm{d}E\,\,  \xi\, n_\mathrm{H}}{\int_{E_\mathrm{l}}^{E_\mathrm{h}} E^3  [\exp{(\frac{E}{k_\mathrm{B}T})}-1]^{-1} \,\mathrm{d}E\,\, 4 \pi^2} \\[6pt]
        &\quad\quad\quad\quad\quad\quad\quad\quad\quad\quad\quad\quad \cdot \sum_{\mu_\mathrm{i},\Phi_\mathrm{e}}  \frac{\mu_\mathrm{i}w_\mathrm{\mu_i} \,\Delta\mu_\mathrm{i}\,S(E)}{N_\mathrm{tot}\,\Delta E \,\Delta\mu_\mathrm{e}} \textrm{ ,}
    \end{split}
\end{equation}
where $\Delta\mu_\mathrm{i} = 0.1$ and $\Delta E$ are the corresponding bin sizes and $w_\mathrm{\mu_i} = \{\frac{1}{3},\frac{4}{3},\frac{2}{3},\frac{4}{3},\frac{2}{3},\frac{4}{3},\frac{2}{3},\frac{4}{3},\frac{2}{3},\frac{4}{3},\frac{1}{3}\}$ are the Simpson coefficients for the 11 values of $\mu_\mathrm{i}$ between 0 and 1. Equation (\ref{norm2}) is Equation (\ref{norm1}) modified for the summation in the $\mu_\mathrm{i}$ and $\Phi_\mathrm{e}$ angles. For the coarse grid chosen, the integration in $\mu_\mathrm{i}$ via the Simpson rule gives more accurate polarization predictions for the isotropic incident intensity, i.e., a physical scenario of a slab illuminated by an extended source above. 

The amplitude change with $\xi$ is observable in the top row of Figure \ref{fig:energy_dep}, as well as the dramatically different flux levels between the thermal and non-thermal reflection due to the ionization parameter definition (\ref{xi}) and the adopted cut-offs. We obtain the same reflection spectral features as described in \citet{Ballantyne2004, Garcia2022}, following up on decades of reflection studies with power-law irradiation \citep[see, e.g.,][]{Fabian2000}. Notably, a low ionization parameter allows photo-electric absorption to dominate at lower energies and, on top of the continuum, we observe the iron line complex near 6--7 keV and a forest of spectral lines below 3 keV. The higher the ionization parameter, the more the reflecting medium acts as a pure electron-scattering mirror to the incident radiation. Although the reflected spectrum from highly ionized matter is spectroscopically nearly indistinguishable from the incident spectrum, it will show distinct polarization features.

The directional Stokes parameters $Q$ and $U$ stored in the individual cells of the local reflection tables typically form $\Psi$ according to the incident and emission beam direction and a dominant corresponding single-scattering angle. For unpolarized incident beam, the emergent polarization angle is nearly perpendicular to the dominant plane of scattering defined by the incident and emission directions. If the associated prevailing single-scattering angle is close to $90^\circ$, then the associated locally emergent polarization fraction is the highest (close to $100\,\%$ for high absorption relative to scattering). It occurs thanks to the Thomson scattering law valid for low energies, although the {\tt STOKES} code makes calculations with all features of true inelastic Compton scattering, which enforces modifications to the simplified description. The predominant polarization angle is constant with energy and averages to $\Psi = 90^\circ$ (negative $p$) for the chosen incident isotropic intensity scenario shown in Figure \ref{fig:energy_dep}. This is consistent with the results of \cite{Poutanen1996b}. Contrary to the isotropic incident flux illumination, for the isotropic incident intensity, we weight by $\mu_\mathrm{i}$ in a way that causes a predominant meridional plane of scattering, resulting in negative polarization. In general, the inelasticity of Compton scattering may alter the polarization sign of the continuum at harder energies, as it was also shown in Figure 6 of \cite{Poutanen1996b}. We refer the reader to the Appendix \ref{cross-validation} for further details. We additionally spot the most prominent spectral lines in positive polarization. 

The fluorescent spectral lines act as depolarizers. But note that, apart from the polarized continuum contribution at line energies, the originally unpolarized fluorescent emission can itself become polarized, as the line photons may scatter before escaping the slab, which is taken into account in the presented simulation. The most prominent fluorescent lines may as well on average switch sign (see this effect near 6.4 keV in Figure \ref{fig:energy_dep}), as compared to the rest of the continuum, due to their different scattering paths before escaping the slab. We refer the reader to the Appendix \ref{cross-validation} for more details.

The photo-electric absorption opacity increases towards the soft X-rays, which reduces the number of scatterings and increases the emergent polarization for lower $\xi$ below $\sim 10\,\textrm{keV}$. We see that even for highly ionized slabs with $\xi = 10000  \,\, \textrm{erg} \,\,\, \textrm{cm} \,\,\, \textrm{s}^{-1}$, the low-temperature blackbody radiation with $k_\mathrm{B}T = 0.2 \, \textrm{keV}$ does not lead to a fully ionized structure, which manifests as a polarization peak near 4 keV in the associated brown solid line of the rightmost bottom panel in Figure \ref{fig:energy_dep}. The highest $\Gamma$ shown in Figure \ref{fig:energy_dep} causes observable absorption peaks in polarization even for $\xi = 10000  \,\, \textrm{erg} \,\,\, \textrm{cm} \,\,\, \textrm{s}^{-1}$, although the power-law incident spectra are less steep in the mid and the hard X-rays than most of the blackbody cases shown. The reason is the overall low power-law flux amplitude in the mid and the hard X-rays compared to the blackbody cases, which is given by the ionization parameter definition (\ref{xi}). Interestingly, for very high blackbody temperatures (for example $k_\mathrm{B}T = 10 \, \textrm{keV}$ shown in light green in Figure \ref{fig:energy_dep}), the polarization absorption features at low energies again strongly appear -- for medium and low $\xi$ even more prominently than for the low $k_\mathrm{B}T$ cases shown. This effect is caused by the already significant depletion of the incident flux at low energies below the spectral peak of the associated Planck curve.
\\ \indent
What remains below the continuum absorption and spectral lines effects is the scattering-induced polarization. It may be observed in the bottom row of Figure \ref{fig:energy_dep} (from left to right panels) that increasing ionization gradually removes the line polarization dips and photo-electric absorption continuum polarization peaks. For the highest slab ionization, we see a generally increasing continuum polarization with energy due to the Compton down-scattering energy shifts, which does not disappear with further increase of ionization of the slab. The same X-ray polarization feature is discussed for the slab thermal transmission problem in \citet{Marra2026}. Within the limits of our simulation, the high-energy incident photons may only down-scatter on cold electrons. Then the soft emergent photons are likely to scatter on average more times than the hard emergent photons. The registered increasing average scattering order then decreases the emergent fraction of polarization, as the net scattering geometry becomes isotropic. The effect is larger for smaller $\delta_i$ (or for isotropic incident intensity weighting rather than for flatter $\mu_i$ distributions) and for steeper incident spectra, but generally saturates below $\sim 1\,\textrm{keV}$ where the slab acts as a Thomson scattering semi-infinite electron atmosphere and the {\tt STOKES} simulation values converge to the Chandrasekhar's law for diffuse elastic reflection \citep[Section 70.3,][]{Chandrasekhar1960}. At very hard X-ray energies for low $k_\mathrm{B}T$, the increase with energy of the polarization degree also saturates, and $p$ starts to decrease with energy, which occurs, however, for an extreme lack of photons, $\sim 20$ orders of magnitude below the flux peak. For incident rays closer to the slab plane, the Compton recoil does not alter the energy dependence of polarization due to average low number of scatterings; or it may even result in depolarization at hard X-rays. We refer to the Appendix \ref{cross-validation} for further details.

For the isotropic incident intensity shown in Figure \ref{fig:energy_dep}, the power-law reflection, and low to moderate $\xi$, the competing continuum absorption and scattering effects are jointly creating an inverted Compton hump in polarization near $\sim 30\,\textrm{keV}$ with decreased number of scatterings at lower energies due to absorption and decreased number of scatterings at higher energies due to Compton down-scattering. The inverted peak is somewhat present also for the blackbody reflection, but largely deformed especially for the steep incident spectra at low $k_\mathrm{B}T$.

\begin{figure} 
\centering
\includegraphics[width=1\linewidth, trim={0.25cm 0.5cm 0.3cm 0.3cm}, clip]
{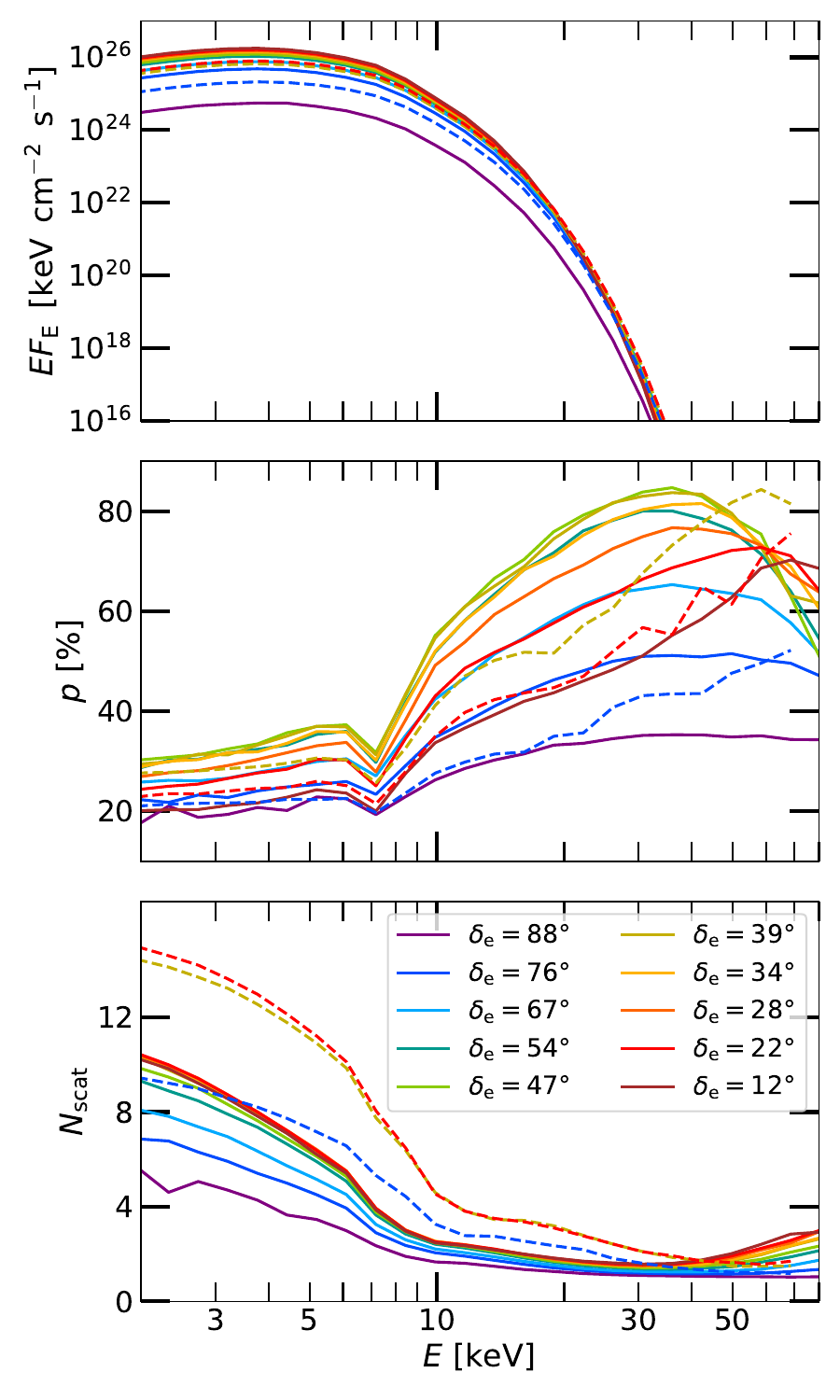}
\caption{Directional local reflection spectra, $EF_\mathrm{E}$ (top), the corresponding polarization degree, $p$ (middle), and the corresponding average number of scattering events, $N_\mathrm{scat}$ (bottom), versus energy in the 2--80 keV band for various emission inclination angles, $\delta_\mathrm{e}$ (in the same color code per panel), one selected incident inclination angle, $\mu_\mathrm{i} = 0.7$, and one selected emission azimuthal angle, $\Phi_\mathrm{e} = 7.5^\circ$. The results are averaged in 7 neighboring energy bins for better MC statistics. The incident spectrum is an unpolarized single-temperature blackbody with $k_\mathrm{B}T = 1\,\textrm{keV}$ and $\xi = 20000  \,\, \textrm{erg} \,\,\, \textrm{cm} \,\,\, \textrm{s}^{-1}$. In solid lines we show the tabulated results, obtained with {\tt STOKES} v2.33. In dashed lines, we show for selected examples $\delta_\mathrm{e} = 22^\circ, 39^\circ, 76^\circ$ the results with the inclusion of Comptonization, obtained with {\tt STOKES} v2.36 in an otherwise identical computational setup.} \label{fig:mue_dep}
\end{figure}

To further demonstrate the link between the scattering-induced polarization energy profile and the energy-dependent average scattering orders, we show in Figure \ref{fig:mue_dep} directional spectra and polarization for one local reflection geometry, i.e., a single {\tt STOKES} simulation with a relatively low $\delta_\mathrm{i}$. {\tt STOKES} allows us to trace the registered average number of scatterings, $N_\mathrm{scat}$, per energy bin. The highest $\xi$ from the computed grid and intermediate $k_\mathrm{B}T$ is displayed, which contains negligible absorption traces and only a small iron-line dip in polarization. We show the spectra in $EF_\mathrm{E} = E^2N_\mathrm{E}$, where the directional photon flux is normalized according to Equation (\ref{norm1}). The results are provided for various emission angles from the slab. It demonstrates that, for reflection, the directional polarization is typically a non-monotonic function of inclination and energy, unlike for the transmission problem, where the polarization degree shows similar energy dependence in the high-energy tail due to down-scattering, but increases monotonically with inclination \citep{Marra2026}. $N_\mathrm{scat}$ is on the contrary a monotonically decreasing function of inclination for reflection, with a broadband decrease with energy due to Compton down-scattering energy shifts, and a slight increase at the very high tail for smaller emission angles due to the Compton-scattering phase function. Above $\sim 30 \,\textrm{keV}$, the reflected photons at high inclinations converge to a single-scattering event. For reflection, the general scattering-induced directional polarization degree profile in inclination and energy is then a combination of the scattering-order changes (in turn dependent on the incident spectral shape), and of the dominant single-scattering angles formed by the incident and emission beams. For the directionality shown in Figure \ref{fig:mue_dep}, the largest polarization degree occurs for a scattering angle of the beam close to $90^\circ$, i.e. for $\delta_\mathrm{e} \sim 45^\circ$.

\section{Discussion}
\label{sec:discussion}

\subsection{The impact of high-energy spectral cut-off}

In the previous section, aside from investigating absorption and spectral-line effects, we interpreted the scattering-induced polarization degree profile with energy as a result of Compton down-scattering effects. Should the down-scattered photons originate in the high-energy tail photon reservoir, the choice of high-energy cut-off of the incident spectra could affect both the reflected spectra and polarization. To test the impact of the high-energy spectral cut-off, we made additional computations for two of the examples shown in the rightmost column \text{($\xi = 10000  \,\, \textrm{erg} \,\,\, \textrm{cm} \,\,\, \textrm{s}^{-1}$)} of Figure \ref{fig:energy_dep}: (i) a single-temperature blackbody reflection with $k_\mathrm{B}T = 5\,\textrm{keV}$, and (ii) a power-law reflection with $\Gamma = 2.2$. Inside the {\tt STOKES} MC simulation, we changed the previously selected value $E_\mathrm{h} = 10^{2.4} \, \textrm{keV}$ by a factor of 0.5 and 2.

Out of the special cases tested, the only non-negligible difference in 2--80 keV in spectra or polarization occurred for the polarization degree for the case of $\Gamma = 2.2$ and the high-energy cut-off that was smaller by a factor of 2. For this case, the reflection spectra were practically identical to the original computation with $E_\mathrm{h} = 10^{2.4} \, \textrm{keV}$, but the polarization in the high-energy tail showed a slightly steeper increase with energy. Thus, we show only this example in a gray dashed line in the rightmost column of Figure \ref{fig:energy_dep} for comparison. The lack of incident photons above 126 keV causes decreased averaged number of scatterings per energy bin for the re-processed photons above $\sim 40\, \textrm{keV}$, resulting in increased polarization fraction that is sensitive to the average scattering order per energy bin. We anticipate from the tested examples that due to a steep decrease of the single-temperature blackbody spectra with energy in the hard X-rays, the 2--80 keV results for the reflected thermal emission shown in the previous section are unaffected by the choice of the high-energy spectral cut-off in the MC simulation.

\subsection{The impact of Comptonization}

In {\tt STOKES} v2.33 used for computation of the main results in this work, Compton down-scattering is taken into account, while Comptonization on hot electrons is neglected. This could potentially be another assumption affecting the generally observed increase of polarization degree with energy in the mid and the hard X-rays due to Compton down-scattering. In the latest development of the {\tt STOKES} code, we can include the scattering effects on hot electrons from the slab temperature structure information obtained consistently from the {\tt TITAN} code. The Comptonization has been implemented into {\tt STOKES} v2.36 during the progress of this work and it is based on the classical frame-to-frame MC approach \citep[e.g.,][]{Krawczynski2012}. It was benchmarked with independent implementation inside {\tt STOKES} with pre-computations on a grid based on the approach in \citet{Poutanen1993} and tested against the {\tt compps} \citep{Poutanen1996} and {\tt MONK} \citep{Zhang2019} codes. The implementation of Comptonization into {\tt STOKES} alongside a detailed investigation of the corrections to the previously published results with {\tt STOKES} that assumed cold electrons is deferred to a future publication. Nonetheless, we are obliged to provide a few special cases additionally computed with the latest version of the code that include the up-scattering effects for comparison.
\\ \indent
In Figure \ref{fig:mue_dep}, we show in dashed lines for three emission angles the directional local reflection spectra, reflected polarization fraction and the corresponding average number of scatterings per energy bin, all corrected for the effects of Comptonization, using {\tt STOKES} v2.36. The results were computed in an otherwise identical computational setup to the previously discussed computations with {\tt STOKES} v2.33 in the same figure. We notice a relatively small decrease in the predicted polarization degree, while the trend of scattering-induced polarization increase with energy remains. For 2--8 keV averaged values in the tested cases, we register a decrease of $p$ from $27.8\, \%$ to $24.2 \,\%$ for $\delta_\mathrm{e} = 22^\circ$, from $33.8 \,\%$ to $28.9 \,\%$ for $\delta_\mathrm{e} = 39^\circ$, and from $24.3 \,\%$ to $21.8\,\%$ for $\delta_\mathrm{e} = 76^\circ$ when up-scatterings are included. The fact that up-scattered photons, to some extent, decrease the polarization degree is related again to the average number of scatterings, which increases in most of the energy range shown when photons are allowed to scatter on hot electrons, as shown in the bottom panel of Figure \ref{fig:mue_dep}. The photons from softer energies than the local electron temperature in the stratified slab get up-scattered, while the high-energy photons still get down-scattered by a non-negligible energy shift, which still creates the prevailing trend in increasing polarization with energy despite the smaller steepness. Consequently, the polarization fraction values above 2 keV presented in this work are overestimated due to the neglect of Comptonization effects. The neglect of Comptonization demands a quantitative correction; however, the arguments presented in the previous section remain qualitatively valid even when up-scattering is included for the range shown.

\subsection{Comparison with {\tt xillverNS}}

The Stokes parameter $I$ from our spectro-polarimetric results can be compared to the {\tt xillverNS} spectra, frequently used for X-ray spectroscopic fitting. In order to do so, we need to re-calculate the {\tt STOKES} normalization according to the different cut-offs used and storage conventions for the {\tt xillverNS} tables. We also integrate the {\tt STOKES} tables in the $\mu_\mathrm{i}$ and $\Phi_\mathrm{e}$ angles, which are implicitly averaged in {\tt xillverNS}. In addition, we average both the {\tt xillverNS} and {\tt STOKES} spectra for emission inclination angles. Then, the raw {\tt STOKES} output for single-temperature blackbody incident spectra needs to be normalized in the following way
\begin{equation}\label{xilnorm}
    \begin{split}
        N_\mathrm{E}(E) & = \dfrac{k_\mathrm{B}T\int_{E_\mathrm{l}}^{E_\mathrm{h}} E^2  [\exp{(\frac{E}{k_\mathrm{B}T})}-1]^{-1} \,\mathrm{d}E\,\,  \xi\, n_\mathrm{H}}{N_{\mu_\mathrm{e}}\int_{E_\mathrm{l}}^{E_\mathrm{h}} E^3  [\exp{(\frac{E}{k_\mathrm{B}T})}-1]^{-1} \,\mathrm{d}E\,\, 4 \pi^2} \\[6pt]
        &\quad\quad\quad\quad\quad\quad\quad\quad\quad\quad\quad \cdot \sum_{\mu_\mathrm{e},\mu_\mathrm{i},\Phi_\mathrm{e}} \dfrac{ \mu_\mathrm{i} w_\mathrm{\mu_i} \,\Delta \mu_\mathrm{i}\,I(E)}{N_\mathrm{tot}\, \Delta E \, \Delta \mu_\mathrm{e}} \textrm{ ,}
    \end{split}
\end{equation}
where $N_{\mu_\mathrm{e}} = 20$ is the number of bins in $\mu_\mathrm{e}$, and \text{$E_\mathrm{l,X} = 0.1 \,\textrm{keV}$} and $E_\mathrm{h,X} = 10^3 \,\textrm{keV}$ are the {\tt xillverNS} cut-offs. The~{\tt xillverNS} spectra need to be additionally multiplied by the $\mu_\mathrm{e}\times 10^{20}/(4\pi)$ factor to consider its storage conventions. In this way, we can compare our results with the {\tt xillverNS} tables for the same combination of the remaining parameters with equivalent normalization.

\begin{figure*} 
\centering
\begin{tikzpicture}[
x=1pt, y=1pt,
inner sep=0pt, outer sep=0pt]
\node[anchor=south west] (base) at (0,0)
  {\includegraphics[width=\linewidth,
                   trim={3.5cm 1.0cm 5.0cm 2.7cm},
                   clip]{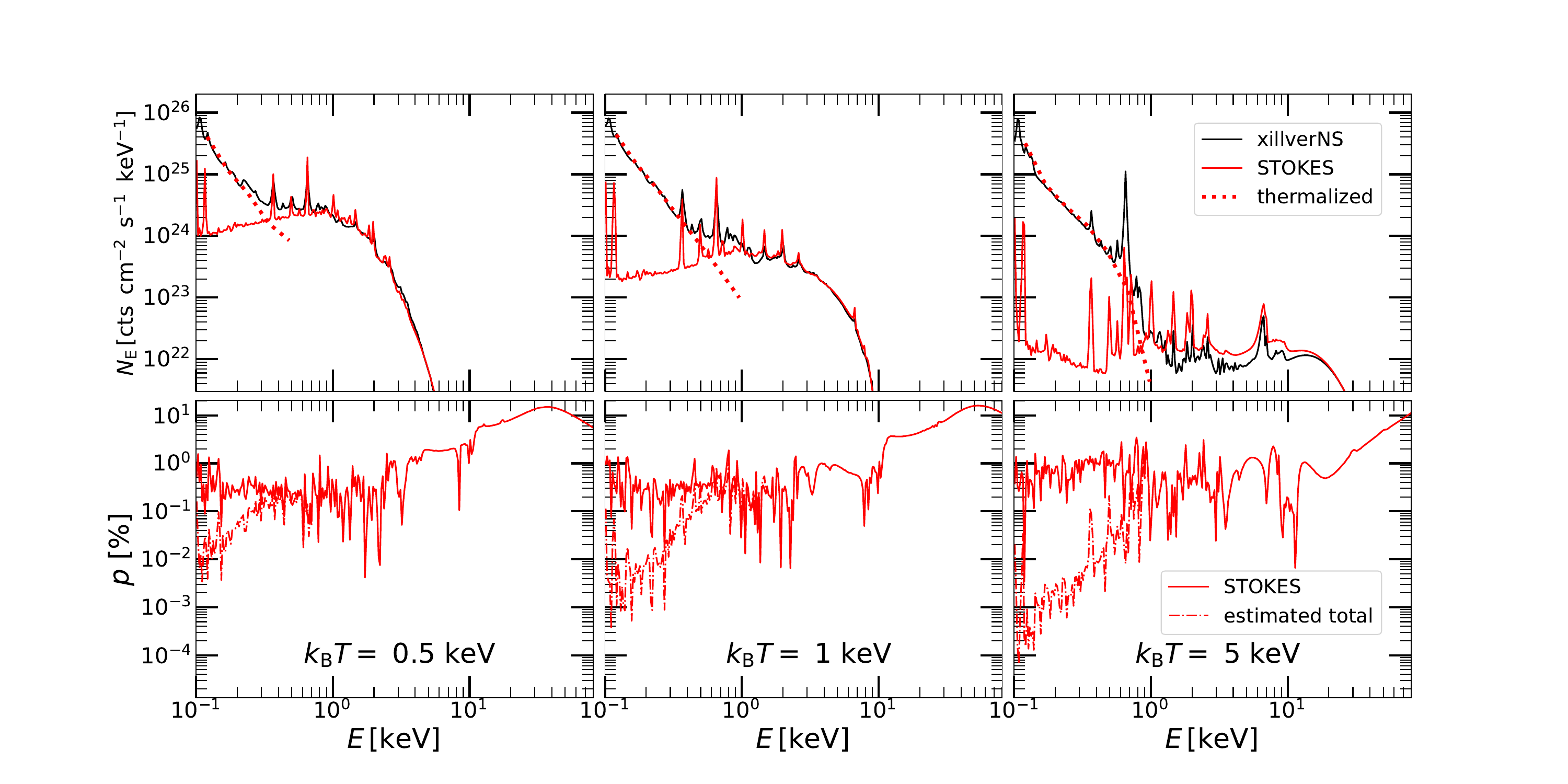}};
\node[anchor=south west] at (40,111)
  {\includegraphics[width=0.11\linewidth]{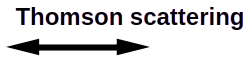}};

\node[anchor=south west] at (34,58)
  {\includegraphics[width=0.20\linewidth]{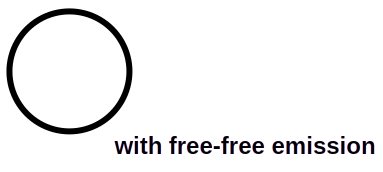}};

\end{tikzpicture}
\caption{Comparison of the presented {\tt STOKES} spectra (in solid red in the top row) and the {\tt xillverNS} spectra (in solid black in the top row) \text{in the 0.1--80 keV band} for identical configurations with $\xi = 1000  \,\, \textrm{erg} \,\,\, \textrm{cm} \,\,\, \textrm{s}^{-1}$ and geometrical averaging over all incident and emission angles. We show three cases of temperature of the incident unpolarized single-temperature blackbody radiation $k_\mathrm{B}T = 0.5,1,5\,\,\textrm{keV}$ from left to right panels, respectively. In the top row, we additionally show the estimated thermalized component contribution (dotted red lines) below 1 keV. In the bottom row, we show the corresponding polarization degree versus energy from {\tt STOKES} (red solid lines) and the same output, corrected for inclusion of the estimated unpolarized thermalized component below 1 keV (dot-and-dashed red lines).} \label{fig:compare_xillver}
\end{figure*}

In the top row of Figure \ref{fig:compare_xillver} we show three comparisons that are reasonable representatives of the entire grid computed, extended down to 0.1 keV. We present the resulting spectra from both calculations for three values of $k_\mathrm{B}T = 0.5,1,5\,\,\textrm{keV}$ and $\xi = 1000  \,\, \textrm{erg} \,\,\, \textrm{cm} \,\,\, \textrm{s}^{-1}$. There is generally a relatively good agreement between the obtained spectral shape above 2 keV, despite the fact that the results were achieved through entirely different methods (an MC simulation and a radiative transfer equation iterative solver). The amplitudes of both reflected spectra are also reasonably in accordance, which verifies the analytically deduced normalization (\ref{xilnorm}) for {\tt STOKES} and its variants (\ref{norm1}) and (\ref{norm2}). The comparisons between the two blackbody reflection spectral results show qualitatively and quantitatively similar differences to those their power-law reflection table counterparts showed when compared, which were discussed in detail in \citet{Podgorny2022}. We refer to this publication for further notes on the comparison between the two methods. We obtain a greater mismatch above 2 keV towards higher incident blackbody temperatures, which is perhaps due to a stronger impact on the amplitude and spectral shape of different cut-offs used for both computations. But, generally, the spectra use different atomic data, different optical depths, and the {\tt XSTAR} code \citep{Kallman2001}, on which the {\tt xillverNS} tables are based, has more than twice more spectral lines implemented in the X-ray band than our model.

It is also worth mentioning that there are many unsolved differences already between the {\tt reflionx} spectra and the {\tt xillver} spectra, and between the {\tt reflionxBB} or {\tt BBrefl} spectra and the {\tt xillverNS} spectra, in some cases reaching an order-of-magnitude different flux values \citep{Garcia2013, Garcia2022}. Moreover, the {\tt xillver} tables have been recently updated for fixes in the {\tt XSTAR} code alongside the atomic data used, which resulted in flux differences by about half an order of magnitude, e.g., near the iron line complex for selected cases \citep{Ding2024}. This underlines the fact that obtaining reflection spectra is generally a difficult task, which relies on extensive computer codes with decades of development. We remind that our tables obtained with the {\tt STOKES} code are not intended for high-precision spectral fitting, but for a reasonable estimate of energy-dependent polarization from reflection off partially ionized slabs in the 2--80 keV range suitable, e.g., for the IXPE or XL-Calibur instruments.

\subsection{The impact of thermalized sub-component}

In the soft X-rays, reflected thermalized photons emerge due to irradiation heating of the slab, which the adopted version of the {\tt STOKES} code itself does not take into account, even though the {\tt TITAN} code computes it consistently. Inside the second, MC simulation part of our calculation, only free-free absorption is consistently included, but not the free-free emission. The thermalized reflection sub-component is generally significant for the single-temperature blackbody spectra impinging on a slab with $n_\mathrm{H} = 10^{15} \, \textrm{cm}^{-3}$ below $\sim 1 \, \textrm{keV}$ \citep{Ballantyne2004, Garcia2022}. For higher densities, the secondary blackbody-like peak of the thermalized sub-component manifests more strongly and can be observable up to $\sim 2\, \textrm{keV}$ \citep{Ballantyne2004, Garcia2022}. It was shown by \citet{Garcia2022} that, above $\sim 2\,\textrm{keV}$, increasing the slab density up to $n_\mathrm{H} = 10^{19} \, \textrm{cm}^{-3}$ to some extent affects the amplitude of the reflected spectra, but to a nearly negligible extent its shape. This means that, regarding our simplified polarization estimates, the 2--80 keV choice for the tabulated range in this work is valid for a larger range of slab densities. For reflection of a power law, the excess of soft impinging photons causes the thermalized component to affect the reflection spectra at even lower energies \citep{Garcia2013,Garcia2016}. In Figure \ref{fig:compare_xillver}, we show an extended range for both {\tt xillverNS} and {\tt STOKES} reflection computations down to 0.1 keV to show the difference in the soft excess due to thermalization. For the spectra in the top row, we show in dotted red lines an estimate of the thermalized sub-component below 1 keV, which comes from subtraction and smoothing of the {\tt xillverNS} and {\tt STOKES} spectra in identical configurations.

Thermally emitted photons are naturally unpolarized. However, the emergent thermalized sub-component may become weakly polarized due to subsequent scatterings of the thermally emitted photons before escaping to the surface of the re-processing layers. The exact polarization fraction will mainly depend on frequency-dependent thermalization depth, slab structure, and emission inclination, with a polarization angle either perpendicular or parallel to the normal to the slab \citep{Nagirner1962,Gnedin1978,LoskutovSobolev1979,Matt1993}. Simplifying the problem and assuming that the thermalized sub-component is unpolarized, we can estimate the total polarization below 1 keV for the selected examples in Figure \ref{fig:compare_xillver}. In the bottom row, we show in solid lines the original polarization obtained from {\tt STOKES}, which includes all major relevant physical processes apart from free-free emission. We obtained the estimated relative flux contribution, $R_\mathrm{T}(E)$, of the thermalized sub-component with respect to the total emission from the spectra below 1 keV shown in Figure \ref{fig:compare_xillver}. From the polarization fraction $p_\mathrm{S}(E)$ obtained by the {\tt STOKES} code with free-free emission switched off, the estimated total reflected polarization fraction, $p_\mathrm{tot}(E)$, can be calculated as $p_\mathrm{tot}(E) = p_\mathrm{S}(E) [1-R_\mathrm{T}(E)]$. The resulting estimate of the total reflected polarization fraction below 1 keV is shown in red dot-and-dashed lines in the bottom panels of Figure \ref{fig:compare_xillver}.
\\ \indent
The relatively low broadband polarization values shown for the selected examples in Figure \ref{fig:compare_xillver} are only a consequence of the choice of an angle-averaged reflection geometry. However, the steep decrease of $p_\mathrm{tot}$ towards soft energies when including a thermalized sub-component, as compared to the relatively constant $p_\mathrm{S}$ with energy in the soft X-rays (as we show examples without significant absorption contribution) from pure {\tt STOKES} simulation, is a more universal result. It can be concluded that, below $\sim 2 \, \textrm{keV}$, depending on the exact location of the secondary thermalization spectral peak, which is in turn mainly dependent on the incident spectrum and the slab density, the polarization decreases rapidly (roughly as $\sim E^{-3}$) towards lower frequencies.

\subsection{The impact of non-constant density and a second thermal source}

All of the aforementioned reflection computations assume a constant vertical density profile of the slab, while, for example, the atmospheres of accretion discs are not expected to have a constant density due to the interplay of gravitational, radiation, and magnetic forces \citep{Blaes2006, Ross2007, Rozanska2011}. Several authors attempted to discuss and compare the reflection spectra for constant density with those computed in a hydrostatic equilibrium with estimated differences in both spectral lines and continuum \citep{Nayakshin2000, Nayakshin2001, Pequignot2001, Ballantyne2001, Rozanska2002, Dumont2002, Ross2007, rozanska2008, Rozanska2011, vincent2016}. Despite the ongoing discussion and difficulties of such efforts, we anticipate that a non-constant density profile modification, which is out of the scope of this work to develop, would result in higher ionization in the layers closest to the surface of the slab due to expansion and heating \citep{Nayakshin2000,Done2007b,Done2010}, which should qualitatively decrease our estimated reflection polarization fraction due to relative absence of absorption opacity in favor of scattering opacity. However, the quantitative difference remains to be computed.

The same direction in correction of the polarization results is expected from the inclusion of a second X-ray thermal source on the other side of the slab,  a natural situation for the photospheres of hot inner XRB accretion discs. The extra source should further ionize the deeper layers of the presented optically thick reflecting slabs \citep{Ross2007, Rozanska2011}, which is again out of the scope of this work to address.

\section{Conclusions}
\label{sec:conclusion}

We calculated the rest-frame X-ray reflection spectra, and polarization for a partially ionized constant-density optically thick slab illuminated by a single-temperature X-ray blackbody radiation. The results in 2--80 keV were converted to a table model, ready to be used inside (relativistic) integrators for particular reflection geometries. Although we plan to address it in a follow-up work, we expect that the main spectro-polarimetric features discovered in the local frame will imprint to the observables at distance from XRB systems. Despite a possible over-prediction of polarization fraction due to the neglect of Comptonization and a secondary thermal source, and the constant density assumption, we argued that the presented results form a foundation for the estimation of mid and hard X-ray polarization of reflected thermal emission from optically thick matter for a large range of densities and ionization states.

Above $\sim 2\, \textrm{keV}$, the reflected thermal emission can be highly polarized up to $100\,\%$, depending on a particular reflection geometry, the incident spectral shape, and the ionization parameter that defines the energy-dependent relative contribution of absorption and spectral lines. Continuum absorption may significantly increase polarization between 2 and 10 keV, where around 6--7 keV, the strong iron-line complex on the contrary depolarizes the re-processed radiation. For highly-ionized slabs, which is also the case for thermally radiating inner disc atmospheres of XRBs, the scattering-induced polarization profile with energy remains. It is typically monotonically increasing with energy due to inelastic Compton scattering, while the slope is dependent on the incident inclination angle or incident angular distribution for extended sources, the shape of the incident spectrum, and the baseline Thomson-scattering energy-independent polarization magnitude valid at lower energies, which is in turn dependent on a particular reflection geometry \citep{Chandrasekhar1960}.
\\ \indent
The obtained rest-frame spectro-polarimetric reflection features are a critical step for the interpretation of the current and future X-ray polarimetric observations of accreting compact objects. Reflection from highly-ionized optically thick matter is a common scenario inside the XRB systems. The more the reflecting medium is ionized, the less spectral features the reflected radiation possesses and it ressembles the incident spectrum, although it remains highly polarized. Hence, it is difficult to estimate spectroscopically a true reflection fraction, but polarimeters may bring insights on hidden reflection geometries. During the first years of IXPE mission operation, many observed accreting compact objects showed relatively high X-ray linear polarization fraction with a monotonic increase with energy in the 2--8 keV band, alongside a constant polarization angle with energy \citep{Dovciak2024,Ursini2024,Marin2024}. Our computational results indicate that, for low impinging rest-frame inclination angles and for steep blackbody spectra, and to a lesser extent for non-thermal spectra, the observed monotonic increase in polarization with energy in the 2--8 keV range could be partly attributed to the intrinsic polarization of highly-ionized reflection component. Such conditions naturally arise for extended thermal sources above an electron-scattering atmosphere, for thermal BH accretion-disc emission returning to the disc due to strong-gravity effects and reflecting under small incident inclination angles close to the BH \citep{Schnittman2009} or for thermal NS accretion-disc emission reflecting from the vertically extended NS boundary/spreading layer \citep{Farinelli2025}.

We note that the independent and simultaneous work to ours, \cite{Farinelli2025}, found increase of polarization fraction with X-ray energy in a mixed-component simulation of a weakly-magnetized accreting NS system when switching off reflection from the accretion disc (their Figure 5f, blue curve). Although definitive proof would require a component decomposition within their simulation, we speculate that this polarization profile as a function of energy is primarily imprinted by the intrinsically energy-dependent polarization of the re-processed thermal bare-disc radiation in the NS boundary/spreading layer, through the mechanisms discussed in our work. The polarization of this single reflection component is diluted by unpolarized direct disc emission with a similar spectral shape, as well as by the weakly polarized, harder Comptonized radiation from the NS, which should mitigate its increase in polarization with energy. On the other hand, the reflection of the thermal radiation from the NS off a highly-ionized accretion disc atmosphere does not provide a strong energy dependence of polarization of the re-processed NS emission, which was also recently shown in another independent MC Comptonization study \cite{Tomaru2026}. This is in line with our results, because we have shown (Figure \ref{fig:ionized_E_dep} without the absorption and spectral-line effects) that for large incident inclination angles (as for the centrally illuminated equatorial disc) the cumulative Compton recoil is not effective.

\cite{Farinelli2025} explain the IXPE data from accreting NS XRBs Cygnus X-2 and GX 9+9 with a multi-component model predicting a sharp polarization increase with energy, neglecting any absorption and spectral-line effects. Despite interpretive complications arising from relativistic and geometrical effects, we suggest that the increase in the total polarization with energy originates not only from the mixing of the highly polarized reflected Comptonized component from the accretion disc with the other three components in their simulation, but also, in part, from the intrinsic increase in polarization with energy of the component reflected from the fully ionized NS boundary/spreading layer. The intrinsic increase in polarization with energy due to inelastic scattering in the disc atmosphere \citep{Marra2026} may also play a non-negligible role in shaping the total polarization profile as a function of energy, although for large electron-scattering optical depths, this component is locally polarized perpendicular to the two reflection components. Polarized reflection from low-ionized or intermediately ionized matter can naturally lead to increasing polarization degree with energy when added to a (less polarized) primary component, because of its spectral shape cut at lower energies due to the typical $\sim E^{-3}$ dependence of photo-electric absorption opacity on X-ray energy.
\\ \indent
Additionally, we estimated the reflection-induced polarization in 0.1--2 keV with added thermalized sub-component due to irradiation of the slab, which was neglected in the code used for the last part of the presented calculations. We expect that the thermalized reflection sub-component will strongly depolarize the reflected continuum, causing the reflected thermal emission in the soft X-rays to be virtually undetectable by the forthcoming soft-X-ray polarimeters, such as the REDSoX instrument \citep{Marshall2018}. For power-law reflection from atmospheres of intermediate ($n_\mathrm{H} \sim 10^{15} \,\textrm{cm}^{-3}$) densities, the depolarization due to bremsstrahlung emission at soft X-rays will likely be shifted to lower energies. But in any case, low or intermediate ionization results in many fluorescent soft-X-ray lines, which generally act as depolarizers.

\begin{acknowledgements}

We would like to thank Bert Vander Meulen for checking the results of our simulations with the {\tt SKIRT} code and providing helpful comments to the manuscript. J.P. and M.D. acknowledge institutional support from RVO:67985815 and thank GACR project 26-22614S. This work was supported by the Ministry of Education, Youth and Sports of the Czech Republic through the e-INFRA CZ (ID:90254). J.P. acknowledges VSB – Technical University of Ostrava, IT4Innovations National Supercomputing Center, Czech Republic, for awarding this project access to the LUMI supercomputer, owned by the EuroHPC Joint Undertaking, hosted by CSC (Finland) and the LUMI consortium through the Ministry of Education, Youth and Sports of the Czech Republic through the e-INFRA CZ (grant ID: 90254). L.M., G.M., and R.T. have been supported by the project PRIN 2022 - 2022LWPEXW - ``An X-ray view of compact objects in polarized light'', CUP C53D23001180006.

\end{acknowledgements}

\bibliography{refs}
\bibliographystyle{yahapj}

\begin{appendix}
\section{{Consistency checks for the fully-ionized and fully-neutral cold-matter limits}}\label{cross-validation}

In this section, we provide several benchmarks of the implementation of down-scattering on cold electrons in {\tt STOKES} v2.33 (used for the main results in this paper) in the X-ray regime, although the code has already been used in the X-rays for spectro-polarimetric predictions \citep[e.g.,][]{Goosmann2011, Marin2012, Marin2018a, Marin2018b, Taverna2021, Podgorny2022, Podgorny2024, Ratheesh2024, Marra2026} with no spotted deviations with respect to similar independent studies. We designed reflection tests, which are additionally illustrative of the scenarios described in this work. We show cases of reflection from fully ionized and fully neutral plane-parallel optically-thick slabs, which were not yet published, despite the extensive testing done on the code. We focus only on power-law reflection (with $\Gamma = 2$), which is, in our opinion, more easily understandable than blackbody reflection and better comparable to existing literature and codes. Plane-parallel geometries have been examined with {\tt STOKES} in transmission for Thomson (i.e., elastic) scattering. The most clear example of cross-validation of the code to classical (semi-)analytical electron-scattering results in transmission is Figure 1 in \cite{Dovciak2008}, which is a reproduction of Figure 6 of \cite{ST1985}, including the Chandrasekhar's polarization prediction for a semi-infinite atmosphere in transmission \citep[Section 68.7,][]{Chandrasekhar1960}. Another direct benchmark for Thomson scattering was performed against the MC code {\tt MONK} \citep{Zhang2019} in Figure 5 of \cite{Zhang2019}.

\begin{figure} 
\centering
\includegraphics[width=1\linewidth,
                   trim={0.45cm 0.5cm 0.4cm 0.35cm},
                   clip]
{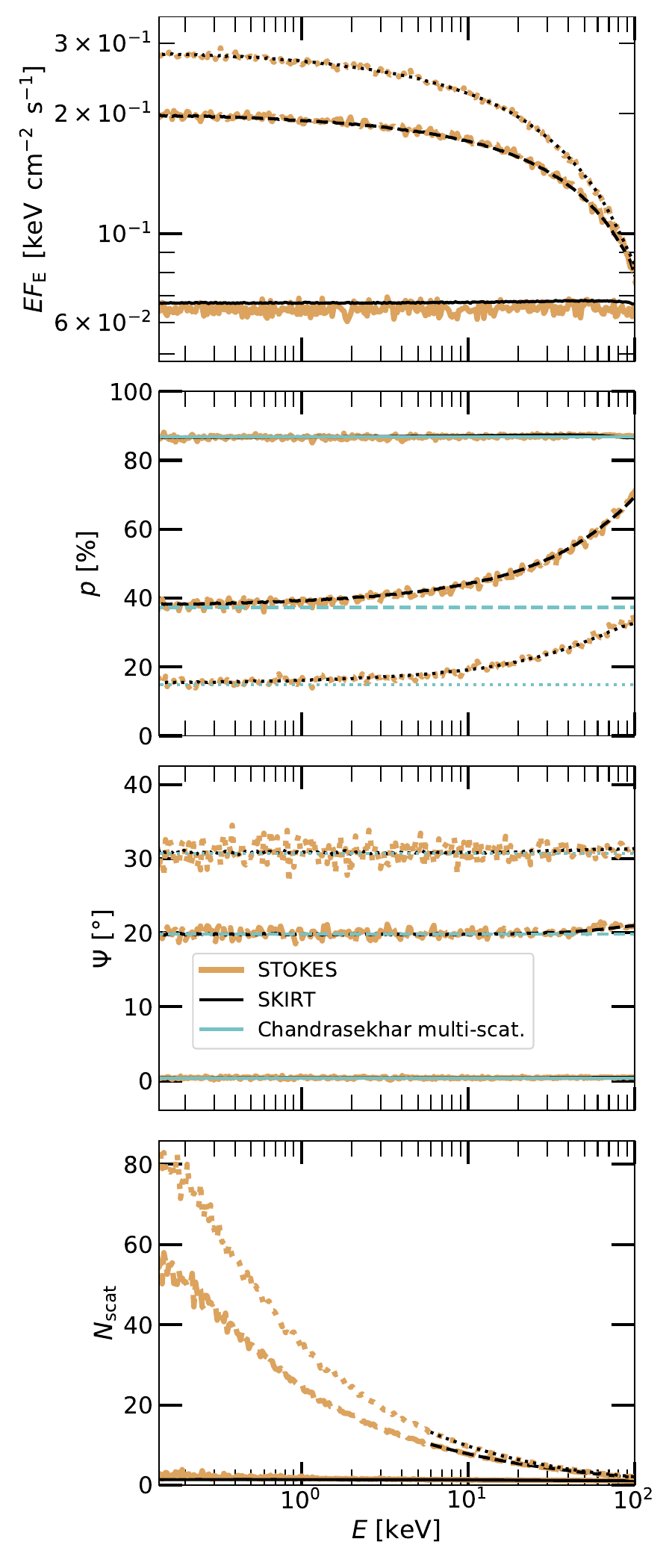}
\caption{The spectra, polarization degree, polarization angle, and the number of scatterings versus X-ray energy reflected from an optically thick electron-scattering atmosphere with the {\tt STOKES} code (orange), {\tt SKIRT} code (black). We also show polarization predictions with Chandrasekhar's multiple-scattering formulae (cyan). We show the results for unidirectional illumination with $\{\mu_\mathrm{i};\mu_\mathrm{e}\}=\{0.025;0.0\}$ (solid lines), $\{0.725;0.3\}$ (dashed lines), $\{0.925;0.7\}$ (dotted lines), all for $\Phi_\mathrm{e} = 76.9^\circ$.} \label{fig:ionized_E_dep}
\end{figure}

Figure \ref{fig:ionized_E_dep} shows the {\tt STOKES} spectra, polarization, and emergent number of scatterings per X-ray energy bin for a reflection from a fully ionized (electrons only) optically thick homogeneous slab of $\tau = 20$. We assume illumination by an unpolarized power law with sharp cut-offs at $E_\mathrm{l} = 10^{-1.1}$ keV and $E_\mathrm{h} = 10^{2.4}$ keV as in the main paper model. We checked that for higher $\tau$ only the $N_\mathrm{scat}$ energy profile slightly changes as photons from the deepest layers contribute significantly in terms of the numbers of scatterings obtained, but the spectra and polarization quantities already converge to the optically thick limit. We show the same quantities obtained with the MC code {\tt SKIRT} v9.0 \citep{Baes2011, VanderMeulen2023, VanderMeulen2024} in identical setup, which provides a useful cross-validation of both independent codes in the X-ray regime and the confirmation of the $N_\mathrm{scat}$ values obtained that are difficult to predict analytically. The absolute values and the dependence on energy of all quantities shown are in a good agreement between the two codes (we normalize the spectra from both codes in the same way, per incident photon in the given range). The solid lines show the illumination nearly along the surface of the slab, which results in a very few ($\sim1$) numbers of scatterings on average. The dashed and dotted lines are, on the contrary, examples when the photon enters deep into the slab and gets essentially trapped before experiencing a large number of scattering events and escaping the slab. We checked that for these two cases, photons that scattered on average even more than 100 times still affect the results below 6 keV and cannot be neglected.\footnote{For this reason we cut the dashed and dotted lines in the bottom panel of Figure \ref{fig:ionized_E_dep} for {\tt SKIRT} at lower energies, because the code cannot provide information on the registered scattering orders beyond 100 without modification of the code. Spectra and polarization predictions are unaffected by this feature of the code.} We additionally show the energy-independent Chandrasekhar's polarization predictions for diffuse reflection \citep[Section 70.3,][]{Chandrasekhar1960} to demonstrate that at soft energies, the results converge to the elastic scattering limit and that the departure at harder observed energies and for large $\mu_\mathrm{i}$ is due to the Compton recoil of individual scattering events on cold electrons, as described in the main paper body.

\begin{figure} 
\centering
\includegraphics[width=1\linewidth,
                   trim={0.45cm 0.5cm 0.4cm 0.4cm},
                   clip]
{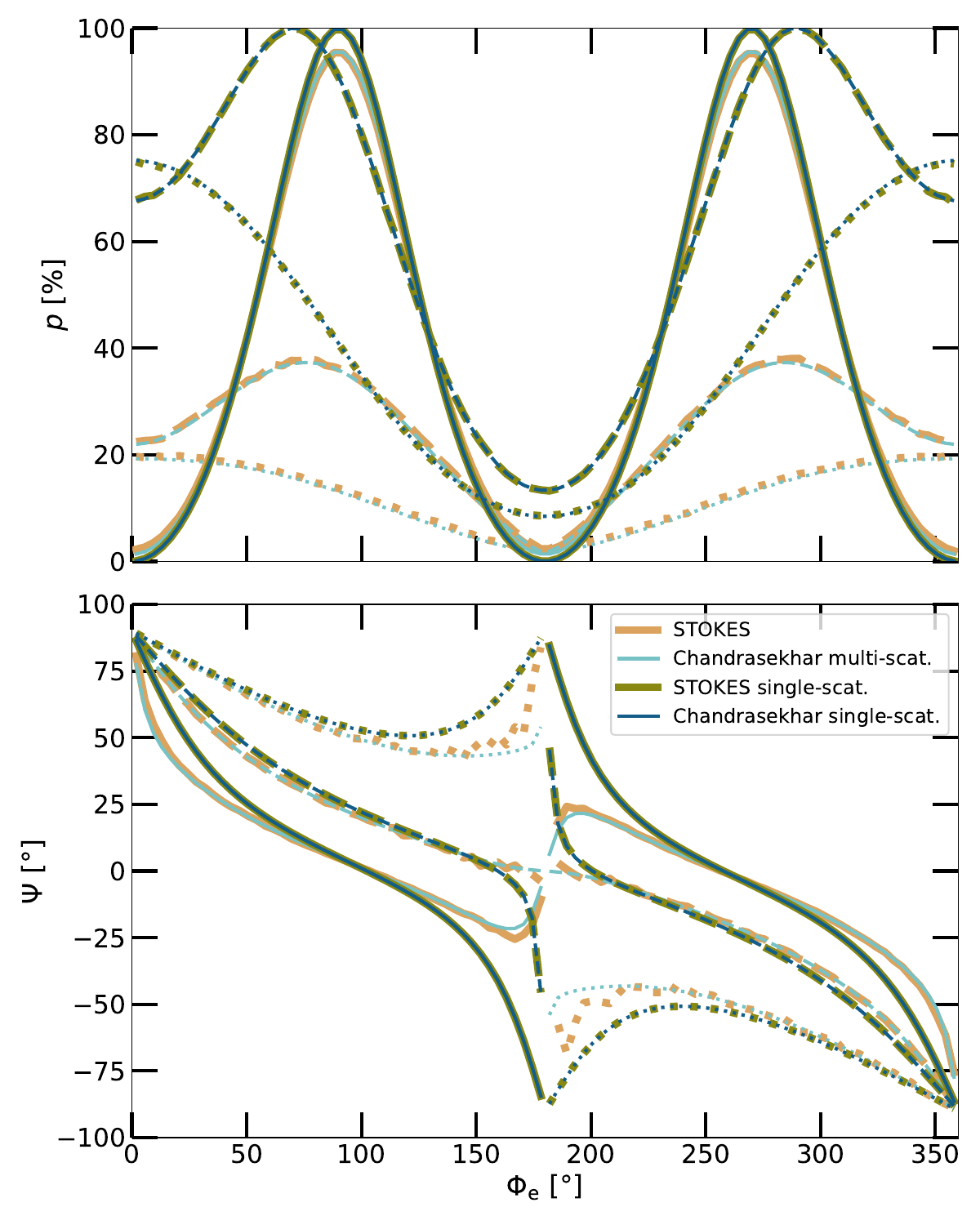}
\caption{Comparison of the polarization results with {\tt STOKES} including/excluding multiple scatterings integrated in 0.15--0.3 keV (orange/green), Chandrasekhar's multiple-/single-scattering formulae (cyan/blue). The top panel shows the polarization degree versus $\Phi_\mathrm{e}$ for the same combinations of $\{\mu_\mathrm{i};\mu_\mathrm{e}\}$ in different line types as in Figure \ref{fig:ionized_E_dep}. The bottom panel shows the polarization angle versus $\Phi_\mathrm{e}$ for $\{\mu_\mathrm{i};\mu_\mathrm{e}\}=\{0.425;0.1\}$ (solid lines), $\{0.525;0.5\}$ (dashed lines), $\{0.725;0.9\}$ (dotted lines). Otherwise, the setup is the same as in Figure \ref{fig:ionized_E_dep}.}\label{fig:vs_Phi}
\end{figure}

Figure \ref{fig:vs_Phi} shows the corresponding plots of polarization degree and angle integrated in 0.15--0.3 keV versus $\Phi_\mathrm{e}$ for different combinations of $\mu_\mathrm{i}$ and $\mu_\mathrm{e}$. We obtained a good match with the polarization predictions by Chandrasekhar's multiple-scattering reflection formulae for elastic scattering due to the low-energy band chosen (see Figure \ref{fig:ionized_E_dep}). The inelasticity of Compton scattering creates a discrepancy, which is also seen in transmission in Figure 1 of \cite{Taverna2021} when comparing the results for $\tau = 20$ with the classical elastic-scattering predictions for semi-infinite atmospheres. In Figure \ref{fig:vs_Phi} we also compare the Chandrasekhar's predictions for single-scattering reflection \citep[Section 70.4,][]{Chandrasekhar1960} with the {\tt STOKES} simulation when only photons that scattered one time are registered in an otherwise identical setup. Apart from a consistency check of the code, this figure illustrates the multiple-scattering effects on polarization mentioned in the main paper body.

\begin{figure*} 
\centering
\includegraphics[width=0.8\linewidth]
{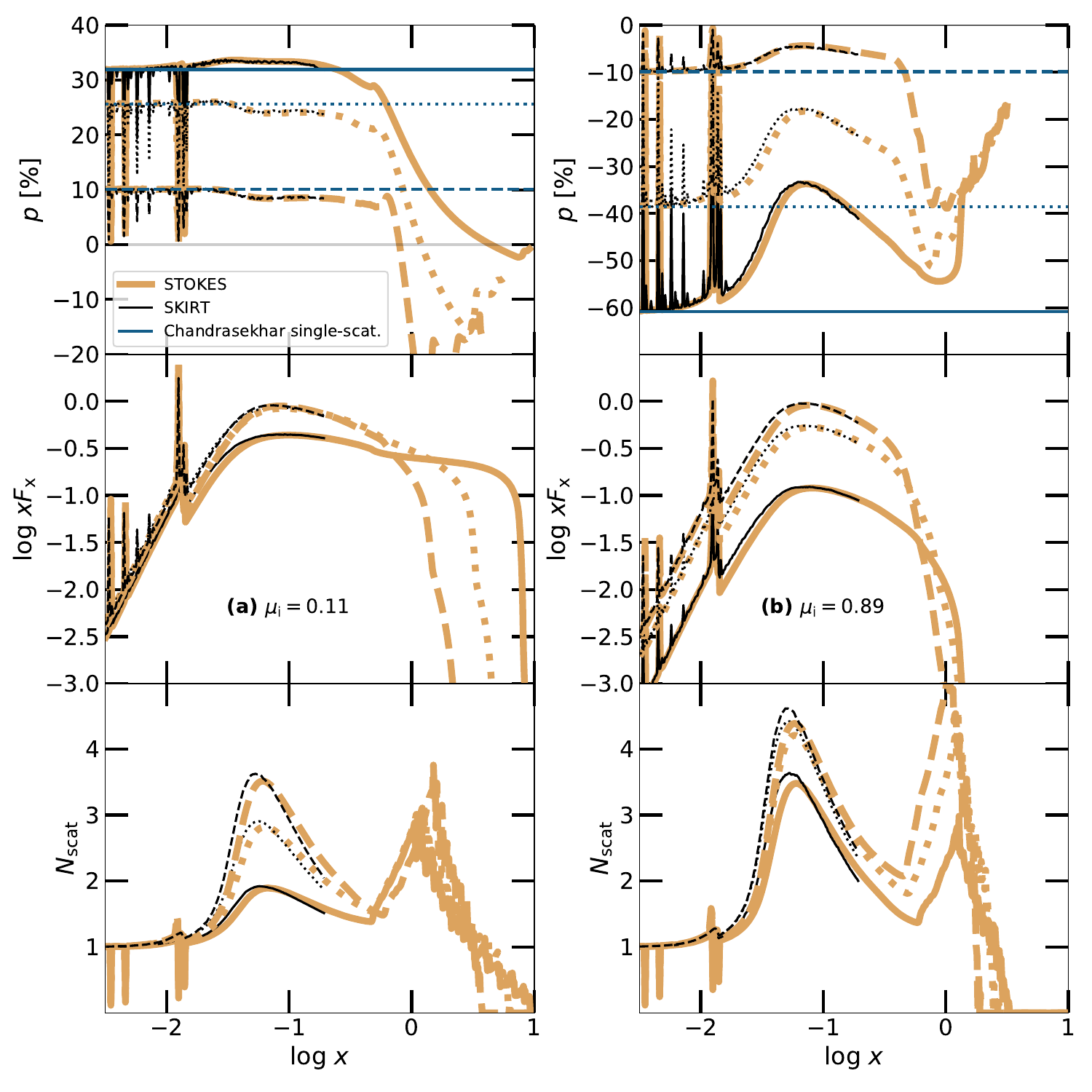}
\caption{Re-creation of Figure 5 from \cite{Poutanen1996b} with the {\tt STOKES} (orange) and {\tt SKIRT} (black) codes, assuming a reflection of unpolarized power law with $\Gamma = 2$ from cold neutral matter. From top to bottom, we show the reflected polarization, spectra, and number of scatterings per X-ray energy. For the polarization, we also show the energy-independent prediction for single-scattering reflection by Chandrasekhar (blue). Left and right panels show different values of $\mu_\mathrm{i}$ for conical illumination. The {\tt SKIRT} results are cut at $x\sim-0.7$, because the code does not operate beyond 500 keV, which could affect the results above $x\sim-0.7$.} \label{fig:neutral_E_dep}
\end{figure*}

In Figure \ref{fig:neutral_E_dep} we show the comparisons between the {\tt STOKES} and {\tt SKIRT} codes in the same setup, but azimuthally integrated, for a fully neutral power-law reflection (including absorption and spectral lines effects), and for $E_\mathrm{h}$ pushed to the limit of the codes\footnote{$E_\mathrm{h} = 5110$ keV for {\tt STOKES} and $E_\mathrm{h} = 500$ keV for {\tt SKIRT}.}. The quantities, ranges, and applied normalization are identical to the Figure 5 in \cite{Poutanen1996b} to show not only a good agreement between the two MC codes, but with the approach described in \cite{Poutanen1996b} that provided exact numerical solution of the radiative transfer equation. We use $x=\frac{E}{m_\mathrm{e}c^2}$ where $m_\mathrm{e}c^2$ is the electron rest energy. We additionally show the $N_\mathrm{scat}(E)$ predictions and agreement for power-law reflection from cold matter with both MC codes in the bottom panel. We also added the energy-independent Chandrasekhar's polarization predictions for elastic single-scattering reflection, suitable for the description of neutral reflection at soft X-rays and at larger wavelengths beyond the X-ray band. The small discrepancies between the two model predictions may be still due to the different levels of complexity of the atomic physics in both codes, and due to the different abundances used for this comparison\footnote{\cite{Asplund2005} for {\tt STOKES} and \cite{Anders1989} for {\tt SKIRT}.}. The different polarization angles obtained in the left and right panels and the significantly smaller $N_\mathrm{scat}(E)$ compared to the fully ionized reflection illustrate the effects discussed in the main paper body.

\begin{figure*} 
\centering
\includegraphics[width=0.8\linewidth]
{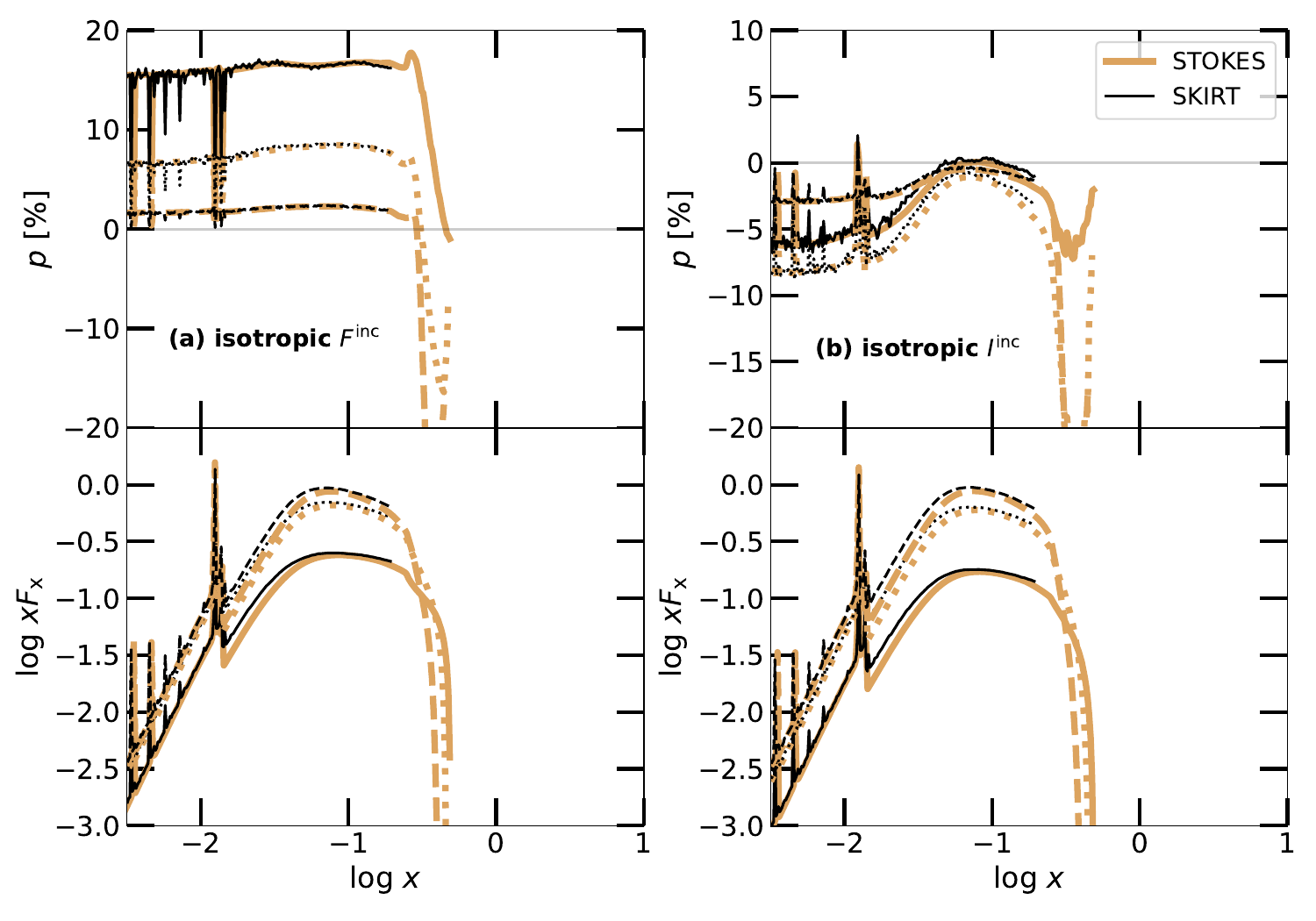}
\caption{Re-creation of Figure 6 from \cite{Poutanen1996b} with the {\tt STOKES} (orange) and {\tt SKIRT} (black) codes, assuming a reflection of unpolarized power law with $\Gamma = 2$ from cold neutral matter with illumination, which is isotropic in incident flux, $F^\mathrm{inc}$, and isotropic incident intensity, $I^\mathrm{inc}$, in the left and right panels, respectively. The reflected polarization (top) and spectra (bottom) are shown in an otherwise identical setup to Figure \ref{fig:neutral_E_dep}.} \label{fig:neutral_E_dep_isotropic}
\end{figure*}

To complete the consistency checks, Figure \ref{fig:neutral_E_dep_isotropic} represents a re-created Figure 6 from \cite{Poutanen1996b}, which is the same as in Figure \ref{fig:neutral_E_dep}, but for isotropic illumination of the slab. Again, good agreement between the MC codes and the radiative-transfer results is reached. The high-energy cut-off used for {\tt STOKES} is now again $E_\mathrm{h} = 10^{2.4}$~keV, because we integrated optically thick fully neutral power-law reflection tables, presented and described in \cite{Podgorny2025}, instead of making new calculations ({\tt STOKES} simulations suffer from orders of magnitude higher computational times as compared to {\tt SKIRT}). Thus, we also used the Simpson coefficients $w_\mathrm{\mu_i}$ for the integration in $\mu_\mathrm{i}$ (see Equation (\ref{norm2})) for the {\tt STOKES} tables used, because these fully neutral power-law reflection tables were calculated for the same 11 bins in $\mu_{\mathrm{i}}$ as the blackbody reflection tables presented in this paper. With {\tt SKIRT}, we used 30 bins in $\mu_\mathrm{i}$ of equal size between 0 and 1, which already ensures convergence to the exact results without the need of the Simpson rule for integration. We note the surprisingly good agreement with the results in \cite{Poutanen1996b} despite their usage of only 3 points in the incident angles. Both MC codes show a positive polarization degree in the most prominent iron spectral lines near 6.4 keV for the isotropic intensity illumination, as opposed to the negative polarization angle obtained in the majority of the energy range due to the predominant meridional scattering planes in such a configuration. The reason is that the fluorescent photons originate on average in the shallow layers and obtain polarization angle parallel to the slab's normal, acting as transmitted photons through optically-thin electron-scattering slabs or plane-parallel emitting atmospheres with a certain type of source functions \citep{Nagirner1962, ST1985}.

\end{appendix}

\end{document}